# Open-Source Tool for Evaluating Human-Generated vs. AI-Generated Medical Notes Using the PDQI-9 Framework

Iyad Sultan, MD[1,2,3]
1 Artificial Intelligence Office, King Hussein Cancer Center, Amman, Jordan
2 Department of Pediatrics, King Hussein Cancer Center, Amman, Jordan
3 Department of Pediatrics, Faculty of Medicine, University of Jordan

Email: isultan@khcc.jo

## Abstract

**Background:** The increasing use of artificial intelligence (AI) in healthcare documentation necessitates robust methods for evaluating the quality of AI-generated medical notes compared to those written by humans. This paper introduces an open-source tool, the Human Notes Evaluator, designed to assess clinical note quality and differentiate between human and AI authorship.

**Methods:** The Human Notes Evaluator is a Flask-based web application implemented on Hugging Face Spaces. It employs the Physician Documentation Quality Instrument (PDQI-9), a validated 9-item rubric, to evaluate notes across dimensions such as accuracy, thoroughness, clarity, and more. The tool allows users to upload clinical notes in CSV format and systematically score each note against the PDQI-9 criteria, as well as assess the perceived origin (human, AI, or undetermined).

**Results:** The Human Notes Evaluator provides a user-friendly interface for standardized note assessment. It outputs comprehensive results, including individual PDQI-9 scores for each criterion, origin assessments, and overall quality metrics. Exportable data facilitates comparative analyses between human and AI-generated notes, identification of quality trends, and areas for documentation improvement. The tool is available online at https://huggingface.co/spaces/iyadsultan/human_evaluator .

**Discussion:** This open-source tool offers a valuable resource for researchers, healthcare professionals, and AI developers to rigorously evaluate and compare the quality of medical notes. By leveraging the PDQI-9 framework, it provides a structured and reliable approach to assess clinical documentation, contributing to the responsible integration of AI in healthcare. The tool's availability on Hugging Face promotes accessibility and collaborative development in the field of AI-driven medical documentation.

## Introduction

The healthcare industry is experiencing a transformative shift with the integration of artificial intelligence (AI) into various clinical and administrative processes. Among these applications, the use of AI for generating medical notes is gaining traction, promising to alleviate the documentation burden on healthcare providers and enhance efficiency. As AI systems become more sophisticated in generating clinical text, it is crucial to establish reliable methods for evaluating the quality and clinical utility of these AI-generated notes in comparison to traditional human-authored documentation.

Evaluating the quality of medical notes is a complex task, requiring consideration of multiple dimensions beyond mere accuracy. High-quality clinical documentation should be up-to-date, accurate, thorough, useful, organized, comprehensible, succinct, synthesized, and internally consistent. To address this multidimensional nature of quality assessment, the Physician Documentation Quality Instrument (PDQI-9) was developed and validated as a comprehensive rubric for evaluating clinical notes across nine key criteria [1]. This instrument has been widely adopted to assess physician-written notes and is increasingly being used to evaluate notes generated by AI systems [2].

In light of the growing need for standardized evaluation tools in the era of AI-driven healthcare documentation, we introduce the Human Notes Evaluator, an open-source tool designed to facilitate the comparative assessment of human-generated and AI-generated medical notes. This tool leverages the established PDQI-9 framework, providing a user-friendly platform for researchers, clinicians, and developers to systematically evaluate and compare the quality of clinical documentation. By making this tool openly available, we aim to promote transparency, collaboration, and rigorous evaluation in the evolving landscape of AI in healthcare.

## Methods

The Human Notes Evaluator is implemented as a Flask-based web application, hosted on Hugging Face Spaces to ensure public accessibility: Human Notes Evaluator. The tool is structured with a frontend built using HTML templates and responsive CSS styling, ensuring usability across different devices (Fig. 1). The backend is powered by a Python Flask application, which manages data processing, session handling, and application logic. Data is stored and managed using CSV files, chosen for their simplicity and portability, with robust parsing capabilities to handle various data formats. The application incorporates comprehensive error handling and logging to ensure reliability and facilitate debugging.

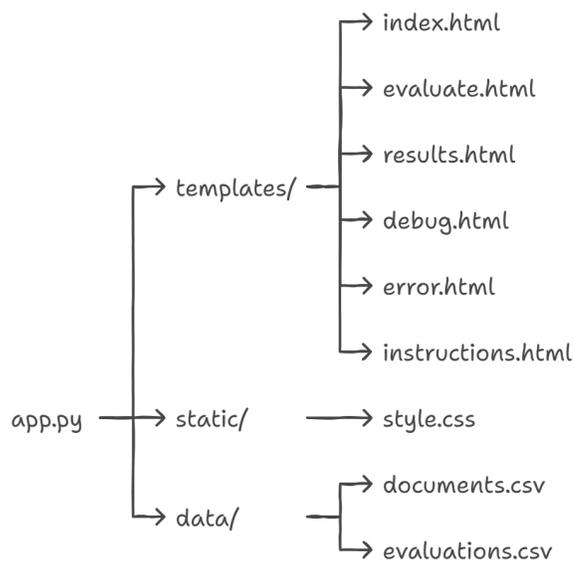

Fig 1. A diagram showing the main components of the human evaluator. Full details are present at https://huggingface.co/spaces/iyadsultan/human_evaluator

At the core of the Human Notes Evaluator is the Physician Documentation Quality Instrument (PDQI-9) [1]. This validated rubric consists of nine criteria, each evaluating a distinct aspect of clinical note quality. The criteria are:

1. **Up-to-date:** Reflects the inclusion of the most recent patient information.
2. **Accurate:** Assesses factual correctness and absence of errors.
3. **Thorough:** Evaluates the completeness of the note in addressing relevant patient issues.
4. **Useful:** Judges the relevance and value of the information for clinical decision-making.
5. **Organized:** Measures the logical structure and arrangement of the note.
6. **Comprehensible:** Assesses the clarity and ease of understanding.
7. **Succinct:** Evaluates conciseness and avoidance of redundancy.
8. **Synthesized:** Determines the integration of information and coherent assessment/plan.
9. **Internally consistent:** Checks for contradictions within the note.

For each clinical note evaluated using the tool, users, typically trained healthcare professionals or researchers, are prompted to rate each of the nine PDQI-9 criteria on a 5-point Likert scale (1 = "not at all", 5 = "extremely"). In addition to the PDQI-9 scoring, the tool includes a feature to assess the perceived origin of the note, with options to classify it as "Human written note," "Generative AI note," or "Unable to determine." This allows for direct comparison of quality scores based on the perceived authorship.

To utilize the Human Notes Evaluator, users prepare a CSV file containing clinical notes with columns for filename, description, medical record number (MRN), and the full note text. The tool's interface guides users through uploading this data, conducting evaluations note by note, and submitting their ratings. Upon completion of the evaluation process, the tool provides an interactive results page, displaying all evaluated documents, PDQI-9 scores for each criterion, origin assessments, evaluator information, and timestamps. This data can be exported in CSV format for further statistical analysis and report generation.

ChatGPT 4.5 was used to write the background for this paper, Claude3.7-sinnet-thinking (agent) was used to generate the code of the evaluator and write the methodology section, while gemini 2.0 flash thinking experimental model was used to write the results and discussion sections, Napkin.ai was used to generate the diagram. All tools were accessed on March 7th, 2025.

## Results

The Human Notes Evaluator (fig. 2,3) is designed to generate comprehensive and exportable results that facilitate in-depth analysis of clinical note quality. The tool's output includes:

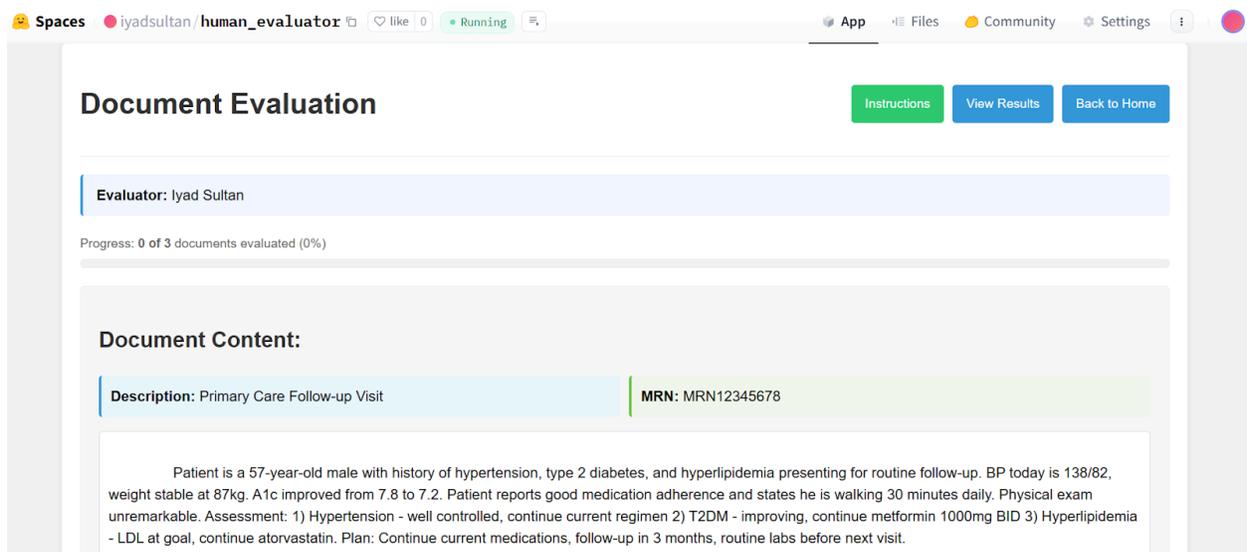

Fig 2. The document is shown with description and MRN for the evaluator to go check the accuracy of source information

Fig 3. Evaluators rank the 9 criteria using a scale from 1 (Not at all) to 5 (Extremely). At the end they are asked whether they know the origin of the document

- **Individual Note Evaluations:** For each note, the results display the scores for each of the nine PDQI-9 criteria, rated on a 1-5 Likert scale. This granular data allows for detailed profiling of note quality across different dimensions.
- **Origin Assessment:** The assessed origin of each note (Human, AI, or Undetermined) is recorded, enabling comparative analysis of PDQI-9 scores based on perceived authorship.

- **Aggregated Quality Metrics:** The tool automatically calculates summary statistics, such as average scores for each PDQI-9 criterion, overall PDQI-9 scores (sum of the nine criteria), and score distributions for human-generated versus AI-generated notes. These metrics provide a high-level overview of the quality differences between note types.
- **Exportable Data:** All evaluation data, including individual note scores, origin assessments, and summary metrics, can be exported in CSV format. This allows users to conduct further statistical analyses using external tools, such as t-tests or ANOVA to formally compare the quality of human and AI-generated notes, or regression analyses to identify factors influencing note quality.
- **Visualizations:** While the current version focuses on data export, future iterations could incorporate built-in visualizations, such as box plots comparing PDQI-9 score distributions between human and AI notes, or radar charts illustrating the strengths and weaknesses of different note types across the nine quality criteria.

These results collectively offer a robust framework for understanding and comparing the quality of clinical notes, facilitating evidence-based assessments of AI in medical documentation.

## Discussion

The Human Notes Evaluator offers a significant contribution to the field of medical informatics and AI in healthcare by providing an open-source, standardized tool for evaluating clinical note quality. By implementing the validated PDQI-9 framework, the tool ensures that evaluations are conducted using established and clinically relevant criteria. This is particularly crucial as AI-generated medical notes become more prevalent, necessitating rigorous methods to ensure their quality and safety.

The tool's user-friendly web interface, deployment on Hugging Face Spaces, and CSV-based data handling make it accessible to a wide range of users, including researchers, clinicians, and AI developers. The exportable data format further enhances its utility, allowing for in-depth statistical analysis and integration with existing research workflows.

Several potential applications and future directions emerge from this work:

- **Comparative Studies:** Researchers can utilize the tool to conduct large-scale comparative studies of human-generated versus AI-generated notes across various clinical settings and AI models. This can provide valuable insights into the current capabilities and limitations of AI in medical documentation.
- **AI Model Development and Refinement:** AI developers can integrate the Human Notes Evaluator into their development pipelines to iteratively evaluate and refine AI models for note generation. The PDQI-9 scores can serve as objective metrics to track progress and identify areas for improvement in AI performance.
- **Clinical Quality Improvement:** Healthcare organizations can use the tool to assess the quality of their existing clinical documentation practices and to evaluate the impact of interventions aimed at improving documentation quality, including the potential integration of AI-assisted documentation tools.
- **Educational Tool:** The Human Notes Evaluator can be used as an educational tool for training medical students and residents on the principles of high-quality clinical documentation and the importance of each PDQI-9 criterion.

Despite its strengths, it is important to acknowledge potential limitations. The PDQI-9 framework, while comprehensive, relies on human evaluation, which can introduce subjectivity. To mitigate this, it is recommended that evaluations are conducted by trained reviewers, and inter-rater reliability measures are employed in research settings. Future developments could explore incorporating automated or semi-automated features to enhance efficiency and reduce subjectivity, while maintaining the clinical relevance of the evaluation process. Furthermore, expanding the tool to incorporate adaptations like the PDSQI-9 [2] to specifically address AI-related issues such as hallucinations and factual omissions would further enhance its utility in the context of AI-generated notes.

In conclusion, the Human Notes Evaluator represents a valuable open-source resource for advancing the rigorous evaluation of medical note quality in the era of AI. By providing a standardized, accessible, and data-driven approach, this tool can contribute to ensuring the quality, safety, and effectiveness of clinical documentation in an increasingly AI-driven healthcare landscape.